\documentclass[conference]{IEEEtran}
\IEEEoverridecommandlockouts
\makeatletter
\newcommand*\titleheader[1]{\gdef\@titleheader{#1}}
\AtBeginDocument{%
  \let\st@red@title\@title
  \def\@title{%
    \bgroup\normalfont\large\centering\@titleheader\par\egroup
    \vskip1.5em\st@red@title}
}
\makeatother
\usepackage{tikz}
\usepackage{cite}
\usepackage{amsmath,amssymb,amsfonts}
\usepackage{algorithmic}
\usepackage{graphicx}
\usepackage{textcomp}
\usepackage{diagbox}
\usepackage{caption,setspace}
\usepackage{algorithmic}
\usepackage{graphicx}
\usepackage{textcomp}
\usepackage{colortbl}
\usepackage{xcolor}
\usepackage{tikz}
\usepackage{diagbox}
\usepackage{enumerate}
\usepackage[figuresright]{rotating}
\usepackage{graphicx}
\usepackage{amssymb}
\usepackage{lineno}
\usepackage{multirow}
\usepackage{colortbl}
\usepackage{hyperref} 
\usepackage{marvosym}
\usepackage{amssymb}
\usepackage{amsfonts}
\usepackage{tikz-qtree}
\def\BibTeX{{\rm B\kern-.05em{\sc i\kern-.025em b}\kern-.08em
    T\kern-.1667em\lower.7ex\hbox{E}\kern-.125emX}}
\title{Segment blockchain: A size reduced storage mechanism for blockchain}
\titleheader{\small https://doi.org/10.1109/ACCESS.2020.2966464}
\begin{document}
\author{\centering
\IEEEauthorblockN{1\textsuperscript{st} Yibin Xu}
\IEEEauthorblockA{\textit{School of Computer Science and Informatics} \\
\textit{Cardiff University}\\
Cardiff, UK \\
work@xuyibin.top}
\and
\IEEEauthorblockN{2\textsuperscript{nd} Yangyu Huang}
\IEEEauthorblockA{\textit{School of Electronic Engineering and Automation} \\
\textit{Guilin University of Electronic Technology}\\
Guilin, China \\
i@hyy0591.me}
}
\maketitle

\begin{abstract}
The exponential growth of the blockchain size has become a major contributing factor that hinders the decentralisation of blockchain and its potential implementations in data-heavy applications. In this paper, we propose segment blockchain, an approach that segmentises blockchain and enables nodes to only store a copy of one blockchain segment. We use \emph{PoW} as a membership threshold to limit the number of nodes taken by an Adversary---the Adversary can only gain at most $n/2$ of nodes in a network of $n$ nodes when it has $50\%$ of the calculation power in the system (the Nakamoto blockchain security threshold). A segment blockchain system fails when an Adversary stores all copies of a segment, because the Adversary can then leave the system, causing a permanent loss of the segment. We theoretically prove that segment blockchain can sustain a $(AD/n)^m$ failure probability when the Adversary has no more than $AD$ number of nodes and every segment is stored by $m$ number of nodes. The storage requirement is mostly shrunken compared to the traditional design and therefore making the blockchain more suitable for data-heavy applications.
\end{abstract}
\section{Introduction}
In the anonymous and autonomous society like a blockchain system, every record should be re-derivable. This feature forms the essential trust of the blockchain, and secures the blockchain; thus, keeping every history transaction is critical. However, the size of the Nakamoto blockchain (Bitcoin \cite{nakamoto2008bitcoin}) has grown from the ground to over $226\ Gbytes$ in the past ten years from January 2009 to June 2019; the size doubled since the February 2017 (at slightly over $100\ Gbytes$) \cite{Bitcoincom}. If the exponential growth to be continued, we are expecting $1.5$ to $1.6\ TBytes$ in Jan 2021, $3$ to $3.2\ TBytes$ in Jan 2022. Parallel to the growth of blockchain size, the storage cost of being a full node (the node which stores all the blocks of the mainchain) in the Bitcoin network is also grown in exponential.

It holds the promise that, through the usage of blockchain, complex and data-demanding jobs can be distributed through predefined protocols (usually through smart contracts \cite {buterin2014next,clack2016smart,savelyev2017contract}) to the anonymous nodes throughout the network, and the majority consensus can secure the job results. Ideally, an alternative-finance system can be built upon---the job publisher pays the system to do tasks, while the anonymous nodes in the system get paid by generating the commonly recognised task results\cite{buterin2014next}. It is guaranteed by the decentralisation and anonymous nature of blockchain that as long as the security threshold---the Adversary not having more than half of the participated calculation power is sustained, the results recognised by the majority are the correct results\cite {nakamoto2008bitcoin}. However, the oversize problem increases the bar of storage requirement for participants, making the system hard to process data-heavy applications like training Artificial Intelligence video recognition models \cite{rota2000activity,samek2017explainable} distributedly and decentralisedly if the system remains universally joinable. Ordinary devices simply do not have enough space to store the data, and the system becomes increasingly centralised if the system process those applications. A transaction in Bitcoin and other Distributed Ledgers \cite{buterin2014next,larimer2014delegated} sized only around several hundreds of bytes. Even with such little usage of data, disadvantaged nodes are gradually leaving the mining game of most Distributed Ledger systems. More and more devices are acting in lightweight mode \cite{dorri2017lsb,gruber2018unifying} or join in mining pools \cite{schrijvers2016incentive,lewenberg2015bitcoin} for the reason of both lacking calculation advantage and the space for storing the blockchain.

Many flavours of approaches like weighted models \cite{dorri2017lsb,gruber2018unifying,khatoonblockchain}, off-chain \cite{poon2016bitcoin,eberhardt2017or}, blockchain sharding \cite{zamani2018rapidchain,kokoris2018omniledger,XUXU} are proposed in recent researches to improve the performance of the blockchain. Many approaches attempt to ease the burden of individual nodes and solve the dilemma among having the ability to process everything, maintaining the decentralised system and increasing the performance. 

For weighted models, the winning chance in the mining game and the duty of a node are different by weights. The lightweight node system \cite{dorri2017lsb,gruber2018unifying} is an example of the weighted model. A lightweight node does not store any block but is the client of some full nodes. They require relevant transactions from the full node to verify a new transaction using Simple Payment Verification (SPV) inquires\cite{guidesimplified}. A lightweight node only takes up to 4.2M Bytes per year, regardless of the total size of blockchain \cite{guidesimplified}, but it cannot verify the new blocks and can be misled by full nodes. In Delegated Proof of Stake (DPoS) \cite {larimer2014delegated}, people elect a fixed number of representatives and contribute their stakes to these representatives;  these  representatives then compete in the game of PoS \cite{vasin2014blackcoin}.  DPoS has excellent performance because the representative nodes usually have a superpower regarding calculation ability, storage, and network bandwidth. These models are now commonly used in many blockchain-powered \emph{IoT} systems \cite{huh2017managing,fan2018roll}, where lightweight nodes are at the edge, or the nodes contribute their stakes to DPoS to function the system. Because the use of authoritarian/superior nodes, the systems are potential-centralised and the system security highly depends on these representatives.

For off-chain approaches, the relevant persons publish a co-signed contract at the beginning and the end of the relationship. Then they do the trading securely through off-chain channels \cite{burchert2018scalable,decker2015fast} without publishing transactions to the blockchain. They only publish the transactions to the blockchain when one violates the off-chain transactions. They need to monitor the blockchain to detect any violations; thus, it is not desirable for users who may go offline. And, there are few usages of off-chain approaches in non-financial related applications. Instead of broadcasting the task and the task result to the network, the entities who use off-chain methods must communicate in private, and this also compromises the anonymous nature of the blockchain. 

For blockchain sharding approaches, they distribute nodes into different shards and divide the storage as well as the jobs to different shards which runs in parallel so that generally the work demand for individual nodes are not increased with the increase of transaction per second globally. Blockchain sharding is designed for applications that require high concurrency and high transaction per second. The design of blockchain sharding approaches mainly focuses on lowering the chance for the Adversary to occupy the majority spots in a shard when the Adversary has taken a relatively large population of nodes but has not taken the majority nodes globally. It is possible for the Adversary not having a security threshold number of nodes globally but controlled a shard, then the security of the system as a whole is compromised. In order to maintain the security of the system, there are very strict requirements over the number of shards and the number of nodes inside a shard \cite{XUXU}. 

We see some blockchain-based storage systems proposed in recent years \cite{xu2018section,xu2018blockchain,wilkinson2014metadisk,vorick2014sia,wang2018forkbase,li2018block,ali2016blockstack,dennis2016temporal,ren2018incentive}, for most cases, the blockchain is only used as the "contract-signing witness" between the \emph{data publisher} and the \emph{data keeper}. Approaches so far are not trying to reduce the size of the blockchain itself and thus cannot avoid the storage pattern of Bitcoin. Not to mention, the nodes not only need to store the transaction (the contract between the \emph{data-publisher} and the \emph{data-keeper}), they also need to keep some data published by the \emph{data-publishers}.

In this paper, we show a new blockchain structure that cut the blockchain into parts. We use blocks as the input to update the states of the \emph{ledger}. Once a block is accepted, the transactions inside are executed, and then a new state is derived basing on the old one. In Bitcoin case, a state is the balance of every wallet address. A fixed number of blocks are placed into a segment; a segment is stored by different nodes and is retrieved by a user only when the user wants to re-derive a state. The nodes keep the latest state so that they can verify the new transactions. They also need to keep a segment assigned by the system to participate in the mining game. In this model, the Adversary may attempt to cause a permanent loss of a specific blockchain segment by storing all the copies of a blockchain segment and then disappear from the network once succeed. The Adversary may also attempt to deceive the system by claiming it stores a part of the blockchain which it does not store.

Segment blockchain dynamically divides the blockchain into $h//s$ segments following the sequence of blockchain. It requires nodes to show a \emph{PoW} (Proof of Work) \cite{bastiaan2015preventing,vukolic2015quest} when joining in the system as well as every time the nodes present the evidence of storing to the system. In this way, we avoid the Sybil attack\cite{douceur2002sybil} and make sure that the Adversary can only take up to $n/2$ of nodes if it has 50\% of the overall calculation power. We use a hypothesis taken from our recent blockchain sharding research \cite{XUXU} to label different nodes in Segment blockchain. We categories the participated nodes into different classes and ensures every part of the blockchain is stored by one node per class. Nodes gain the reward for keeping the blockchain segments in the edited game of mining.

Instead of using that hypothesis to build a blockchain sharding system that is more focusing on the improvement of transaction per second, Segment blockchain concentrates solely on the reduce of blockchain size. In blockchain sharding systems, the honest nodes must be the majority of every shard to keep the security of the system as a whole. However, considering storage, when an Adversary fails to become the keeper of every copy of a Segment, its attack does not succeed. So that, in segment blockchain, we do not require the honest people to be the majority of the people who store a specific segment, we only need to ensure that every segment gets at least a reliable keeper. With the loose security threshold, we can assign less number of nodes to store a segment in order to keep that segment securely, compared to blockchain sharding systems where they need to secure the majority nodes are honest. That is why the storage can be much shrunken. 

Segment blockchain is suitable for most blockchain applications nowadays including notation\cite{lopez2017caterpillar}, identity control \cite{dupont2017blockchain}, multinational customs record exchange \cite {okazaki2018unveiling} which applications do not require a large transaction throughput (like ten thousand to 1 million per second), but get benefit from decentralisation. It also helps the implementation of blockchain in \emph{IoT} environment where the edge devices are lacking storage capacity to keep the full record; meanwhile, the systems are not requiring significant transaction per second \cite{ren2019secure}. Segment blockchain can even be used to improve the blockchain sharding by separating transaction storage from transaction verification: the nodes in shards only keep the latest state in their shards, and the history transactions are placed into Segments and being handled separately by Segment blockchain.

In the following sections, we will show the Segment blockchain in detail, discuss why our method is secured and can provide a $(AD/n)^m$ failure probability; also, how the model prevents the spoof of storing. We will also show the data requirement compared to the traditional Nakamoto blockchain (Bitcoin).

\section{The \emph{Jury} Hypothesis and its usage in Segment blockchain}
We proposed the \emph{Jury} Hypothesis as an analogy of an $n/2$ Byzantine-node tolerate blockchain sharding approach in \cite{XUXU}, which turns the blockchain into multiple committees that run in parallel.

The \emph{Jury} Hypothesis states that the member of the Jury of a court comes from the diverse background, so that when a verdict is reached, it can be seen as the decision reached from the whole society (every class of people). If it takes $m$ different occupations to form a jury, then when there are a $s$ number of court hearings run in parallel, there are $s$ number of people in each one of the $m$ occupation. Table \ref{fig:img3} shows a court schedule; each court represents a shard, $A$ is a person controlled by the Adversary while $H$ is an honest person.

\begin{table}[h!]
\small
\centering
\caption {Court Schedule}
\begin{tabular}{ccccc}
\diagbox{Ocp}{Court number}&0&1&2&3\\
Occupation 1&A&A&A&A\\
Occupation 2&H&A&H&A\\
Occupation 3&A&H&A&H\\
Occupation 4&H&A&H&H\\
Occupation 5&H&H&H&H\\
\end{tabular}
\label{fig:img3}
\end{table}

It is ruled that a verdict is reached when a pre-defined $T$, $T>0.5m$ number of people inside the jury reached a consensus. Assuming there exists a random assignment scheme that assigns people of the same occupation to different courtrooms where different court hearings are taken place in parallel. Then, the chance for the Adversary to gain $T$ spots inside the target courtroom is (assuming the Adversary put all its nodes into the front $T$ occupations)
\begin{equation}
        Pr[T]=\prod_{i=1}^{T}\frac{A_i}{s}
\end{equation}
where $A_i$ is the number of people inside courtroom $i$ who are controlled by the Adversary. To derive the maximised $Pr[T]$, we want $\prod_{i=1}^{T}A_i$ to be maximised because $s$ is the same. Let the Adversary has $AD$ number of people inside the system (Court Jury Schedule), then $AD=\sum_{i=1}^m A_i$. To maximise the value of $\prod_{i=1}^{T}A_i$, we consider 
\begin{equation}
    A_i=\lfloor AD/T \rfloor, i \in [1,T-1]
\end{equation}
\begin{equation}
    A_T=\lfloor AD/T \rfloor+ AD\ mod\ T
\end{equation}
This scenario is the maximised because, given any positive integer $X$, 
\begin{equation}
    X*X > (X-1)*(X+1)=X*X-1
\end{equation}

Thus, 
\begin{equation}
    Pr[T]_{max}\approx(\frac{AD}{T*s})^T
\end{equation}
\subsection{The \emph{Jury} Hypothesis for Segment blockchain}
Let the jury of a court stores part of the blockchain, then for the Adversary to control all the people ($T=m$) inside this jury is 
\begin{equation}
    Pr[T]_{max}\approx(\frac{AD}{m*s})^m
\end{equation}

Let the court system has a $n$ number of jury members in total ($n=m*s$). Then,
\begin{equation}
Pr[m]_{max}\approx(\frac{AD}{n})^m
\end{equation}

If the Adverary has no more than $50\%$ fraction of the people inside the system (Nakamoto blockchain threshold), then \begin{equation}
Pr[T=m]_{max}\approx(\frac{1}{2})^m
\end{equation}
Thus, the maximum chance for a failure to occur is $(\frac{1}{2})^m$.
\subsection{Challenge}
The challenge of implementing \emph{Jury} Hypothesis for the blockchain storage is (1) how to give different nodes different occupations. (2) how to randomly assign storage to a node. (3) how to prove that a node stores a data. (4) how to adjust the membership and reform the jury when nodes go offline.
\section{segment blockchain}
Segment blockchain cuts the blockchain into segments. The size and the number of blockchain segments are dynamically adjusted base on the quantity and the occupation of nodes in the system. Every node only stores one blockchain segment and the block header of every block in the mainchain.
\subsection{Block as input}
\begin{figure}[h!]
\centering
\begin{tikzpicture}[scale=0.9]
\node[rectangle,draw,minimum width=1.1cm, minimum height=1cm,dashed] (a0)at (-1.5,0){\small State 0};
\node[rectangle,draw,minimum width=1.1cm, minimum height=1cm] (a1)at (0,0){\small Block 1};
\node[rectangle,draw,minimum width=1.1cm, minimum height=1cm,dashed] (a2)at (0,-1.5){\small State 1};
\node[rectangle,draw,minimum width=1.1cm, minimum height=1cm] (a3)at (1.5,-1.5){\small Block 2};
\node[rectangle,draw,minimum width=1.1cm, minimum height=1cm,dashed] (a4)at (1.5,-3){\small State 2};
\node[rectangle,draw,minimum width=1.1cm, minimum height=1cm] (a5)at (3,-3){\small Block 3};
\node[rectangle,draw,minimum width=1.1cm, minimum height=1cm,dashed] (a6)at (3,-4.5){\small State 3};
\node[rectangle,draw,dashed,minimum width=0.5cm, minimum height=0.3cm] (a7)at (-3,-4){ };
\node[rectangle,draw,minimum width=0.5cm, minimum height=0.3cm] (a7)at (-3,-4.6){ };
\node(a7)at (-1.5,-4.1){\small Self generable};
\node(a7)at (-.9,-4.7){\small Globally sync required};
\draw [->] (a0) --(a1);
\draw [->] (a1) --(a2);
\draw[->] (a2) --(a3);
\draw[->]  (a3) -- (a4);
\draw[->] (a4) ->(a5);
\draw[->]  (a5) -- (a6);
\end{tikzpicture}
\caption{An example of the \emph{block} and \emph{state}}
\label{fig:img1}
\end{figure}
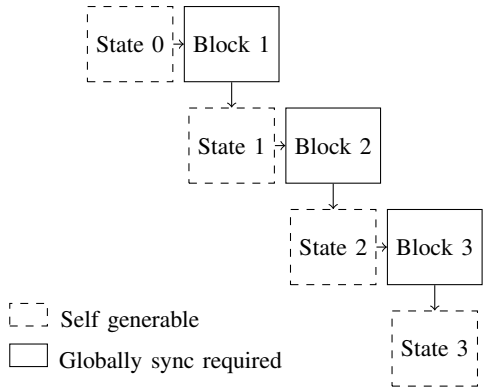

Let the blockchain be a \emph{state} machine where every block is the input of the current \emph{state}; the machine reaches the next \emph{state} after processing the current block. In Segment blockchain, every node keeps the latest \emph{state} and all the block headers while storing some copies of the previous blocks. The blockchain is secured when a node can download all the blocks, run them as \emph{inputs}, and derive the same \emph{state}. Figure \ref{fig:img1} shows an example of this design.

In Bitcoin case, the \emph{state} can be a ledger that records the balance of every account.
\subsection{blockchain segments}
Let there be $s$ number of blockchain segments; every segment takes $h//s$ number of nodes following the index number of the blocks and a \emph{state} derived from the latest block of the previous segment. $h$ is the current length of the blockchain. Except the last segment, other segments are of equal length. The last segment would additionally contain $h \mod s$ blocks. Figure \ref{fig:img2} shows an example of this design.

\begin{table}[h!]
\begin{tabular}{c|c}
\begin{tikzpicture}[scale=0.80]
\node[] (a11) at (0,.3){blockchain segment 1};
\node[rectangle,draw,fill=red, minimum width=3.5cm, minimum height=0.7cm] (a11) at (0,1){ };
\node[rectangle,draw,fill=red, minimum width=3.5cm, minimum height=0.7cm] (a11) at (0,-.5){ };
\node[] (a11) at (0,-1.2){blockchain segment 2};
\node[rectangle,draw,fill=white, minimum width=1.1cm, minimum height=0.5cm] (a1)at (-1.5,1){\small State 0};
\node[rectangle,draw,fill=white, minimum width=1.1cm, minimum height=0.5cm] (a1)at (0,1){\small Block 1};
\node[rectangle,draw,fill=white, minimum width=1.1cm, minimum height=0.5cm] (a3)at (1.5,1){\small Block 2};
\node[rectangle,draw,fill=white, minimum width=1.1cm, minimum height=0.5cm] (a1)at (-1.5,-.5){\small State 2};
\node[rectangle,draw,fill=white, minimum width=1.1cm, minimum height=0.5cm] (a1)at (0,-.5){\small Block 3};
\node[rectangle,draw,fill=white, minimum width=1.1cm, minimum height=0.5cm] (a3)at (1.5,-.5){\small Block 4};
\end{tikzpicture}&
\begin{tikzpicture}[scale=0.90]
\node[] (a11) at (0.65,-.7){blockchain segment 1};
\node[rectangle,draw,fill=red, minimum width=2.2cm, minimum height=0.7cm] (a11) at (0.65,0){ };
\node[rectangle,draw,fill=red, minimum width=2.2cm, minimum height=0.7cm] (a11) at (0.65,-1.5){ };
\node[rectangle,draw,fill=red, minimum width=3.7cm, minimum height=0.7cm] (a11) at (0.65,-2.8){ };
\node[] (a11) at (0.65,-2.2){blockchain segment 2};
\node[rectangle,draw,fill=white, minimum width=1.1cm, minimum height=0.5cm] (a1)at (-0.1,0){\small State 0};
\node[rectangle,draw,fill=white, minimum width=1.1cm, minimum height=0.5cm] (a2)at (1.4,0){\small Block 1};
\node[rectangle,draw,fill=white, minimum width=1.1cm, minimum height=0.5cm] (a3)at (-0.1,-1.5){\small State 1};
\node[rectangle,draw,fill=white, minimum width=1.1cm, minimum height=0.5cm] (a3)at (1.4,-1.5){\small Block 2};
\node[rectangle,draw,fill=white, minimum width=1.1cm, minimum height=0.5cm] (a5)at (-0.9,-2.8){\small State 2};
\node[rectangle,draw,fill=white, minimum width=1.1cm, minimum height=0.5cm] (a7)at (0.5,-2.8){\small Block 3};
\node[rectangle,draw,fill=white, minimum width=1.1cm, minimum height=0.5cm] (a7)at (2,-2.8){\small Block 4};
\node[] (a11) at (0.65,-3.5){blockchain segment 3};
\end{tikzpicture}\\
$h=4$, $s=2$.&$h=4$, $s=3$.\\	
\end{tabular}
\captionof{figure}{The example of the blockchain segments}
	\label{fig:img2}
\end{table}
\subsection{Node Membership}
Let every block has a section which records the pending nodes. The pending nodes are nodes which reported to the system but has not yet been assigned with storage. When a node wants to join in the system, it checks the pending nodes information of the latest block and finds an occupation that is less affluence in number. It then claims that occupation. Let there be a threshold \emph{PoW} difficulty $P$. The node needs to present a \emph{PoW} of a $P\times s$ difficulty to the system before this node's info (the occupation and a public identity key) is written into the pending node section. Table \ref{fig:img222222222} shows the structure of this \emph{PoW}. Until the node has been instructed to store a specific blockchain segment, the node needs to additionally present a $P$ difficulty \emph{PoW} (of the same content) in every iteration of the mining game to the system to keep the spot in the pending node section.

\begin{table}[h!]
\centering
\caption {PoW for pending node}\resizebox{\columnwidth}{!}{%
\begin{tabular}{ccl}
\hline 
Name&Storage&Description\\\hline
\emph{Hash\_Prev\_block}&32 bytes&The hash of the \\
&&block in block height\\
&&\emph{h-s}.\\
\emph{Occupation}&4 bytes& The occupation claimed\\
\emph{Identity\ Key}&32 bytes&Node's public identity key\\
\emph{Nonce}&32 bytes&The number tried for this \emph{PoW}\\
&&to reach the required difficulty\\
\hline
\end{tabular}}
\label{fig:img222222222}
\end{table}

Rank the nodes' info by the time their info is written into the pending section in ascending order into a list $PN$. Let $PN_{i,j}$ refers to the index $j$ pending node of the occupation $i$. Every time when $min(len(PN_{i}))>=10, i \in [1,m]$, the storage of all nodes is re-assigned while the size of the blockchain segment is readjusted and $10$ more blockchain segments are created $(s=s+10)$ and $PN_{i,1..10}, i \in [1,m]$ are added to the system. Table \ref{fig:img33333} and \ref{fig:img333433} shows an example of the nodes in the system and the $PN$ list; Table \ref{fig:img33334} and \ref{fig:img333434} show a possible situation of the nodes in the system and the $PN$ list when the front $10$ elements of every occupation in the $PN$ list (Table \ref{fig:img333433}) are added to the system. Because there is a queue for the pending nodes in every occupation, it is nature for the nodes to claim an occupation that is less affluence in number to join in the system quicker. In this way, we solve the challenge (1) of segment blockchain.
\begin{table}[h!]
\small
\centering
\caption {Nodes in the system}%
\begin{tabular}{ccccccccccc}
\diagbox{Occupation}{blockchain segment}&1&2&3&4\\
Occupation 1&A&A&A&A\\
Occupation 2&H&A&H&A\\
Occupation 3&A&H&A&H\\
Occupation 4&H&A&H&H\\
Occupation 5&H&H&H&H\\
\end{tabular}
\label{fig:img33333}
\end{table}
\begin{table}[h!]
\small
\centering
\caption {PN list}\resizebox{\columnwidth}{!}{%
\begin{tabular}{cccccccccccccc}
\multicolumn{12}{c}{\emph{PN}}\\
Occupation 1&A&A&H&H&A&A&H&H&H&H&H\\
Occupation 2&H&A&A&A&H&H&A&A&A&H&H\\
Occupation 3&A&H&H&A&A&A&H&H&A&A&\\
Occupation 4&H&A&A&H&H&A&H&A&A&H&H\\
Occupation 5&H&H&A&H&A&A&H&H&A&H&H\\
\end{tabular}}
\label{fig:img333433}
\end{table}

\begin{table}[h!]
\centering
\caption {Nodes in the system after adding}\resizebox{\columnwidth}{!}{%
\begin{tabular}{cccccccccccccccccccccc}
\diagbox{Ocp}{Bs}&1&2&3&4&5&6&7&8&9&10&11&12&13&14\\
Occupation 1&H&A&H&H&A&H&A&H&A&A&A&A&A&H\\
Occupation 2&H&A&H&A&A&A&H&H&A&A&H&A&H&A\\
Occupation 3&H&A&A&H&H&A&A&A&H&A&H&A&A&H\\
Occupation 4&A&A&H&A&H&H&A&H&H&H&A&H&H&A\\
Occupation 5&H&A&H&H&H&H&H&H&H&H&H&A&A&A\\
\end{tabular}}
\label{fig:img33334}
\end{table}
\begin{table}[h!]
\small
\centering
\caption {PN list after adding}
\begin{tabular}{cccccccccccccc}
\multicolumn{2}{c}{\emph{PN}}\\
Occupation 1&H\\
Occupation 2&H\\
Occupation 3&\\
Occupation 4&H\\
Occupation 5&H\\
\end{tabular}
\label{fig:img333434}
\end{table}

\subsection{Storage assignment}
When new nodes are added to the system right after block height $h$, let $ID_{i,j}$ refers to the identity key of the node of occupation $i$ which stores $j$ blockchain segment. Create a \begin{equation}
    RID_{i,j}=ID_{i,j}\ hash\ BH_h
\end{equation}
link $RID_{i,j}$ with $ID_{i,j}$ and rank $RID_{i},\ {i} \in [1,m]$ by the ascending order, then adjust $ID_{i}$ according to the sequence of the ranked $RID_{i}$. $BH_i$ is the hash of the block $i$, $hash$ is a hash function that returns an $256\ bits$ integer. After the procedure above, the assignment is completed. In this way, we solved the challenge (2) of segment blockchain.
\subsection{Proof of Storage}
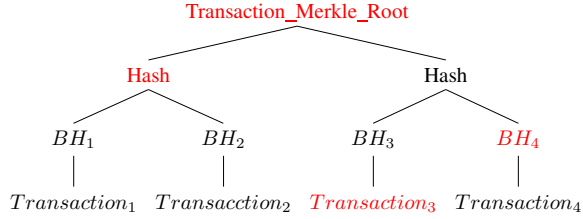
\begin{figure}
\centering
\begin{tikzpicture}[scale=0.80]
\centering
\Tree [.\color{red}{Transaction\_Merkle\_Root} [.\color{red}{Hash} [.$BH_1$ $Transaction_1$ ] [.$BH_2$ $Transacction_2$ ] ]
[ .Hash [.$BH_3$ \color{red}{$Transaction_3$} ]
[.\color{red}{$BH_4$} $Transaction_4$ ] ]] 
\end{tikzpicture}
\caption{The example of a transaction Merkle tree of block $i$. Let $len(i)=4,\ CI_{i}=3$. The Merkle branch for Proof of Storage contains all the nodes in red.}
\label{fig:img2222}
\end{figure}

Let every block header records the Merkle root of the transactions embedded in the block. Let $BH_{h}$ be the block header hash of the latest block height (block height $h$). If the node $j$ in occupation $i$ stored the blockchain segment $k$, then let 
\begin{equation}
    CI_{k}=(BH_{h}\ hash\ ID_{i,j}\ hash\ i)\ mod\ len(k)+1
\end{equation}
where $CI_k$ is the index number of a transaction in blockchain segment $k$, $len(k)$ is a function that returns the number of transactions inside the blockchain segment $k$. When the nodes present the evidence of storing the blockchain segment $k$ at the block height $h$ (referred to as \emph{Proof\ of\ Storage}), it should provide
\begin{itemize}
\item The transaction which $CI_k$ refers to. 
\item A Merkle branch that can derive the Merkle root indicated in the block header of a block $B$ in the blockchain segment $k$.
\item The transaction which $CI_k$ refers to is inside block $B$.
\end{itemize}
Figure \ref{fig:img2222} shows an example of the \emph{Proof\ of\ Storage}. In this way, we solved the challenge (3) of segment blockchain.
\subsection{Mining and blockchain segment size adjustment}
Segment blockchain runs the same mining rule as Nakamoto blockchain for the block creation. Nakamoto blockchain rules that the node which presents a valid block with the most difficult \emph{PoW} in an iteration wins that iteration of the mining game.

Let there be $s$ number of blockchain segments existing in the system, and the current block height is $h$. Then, the nodes who were assigned to store the number $k=(h\ mod\ s)+1$ blockchain segment should send the \emph{Proof\ of\ Storage} to the network after the block of the block height $h$ is created. They need to submit a \emph{PoW} of $P\times s$ difficulty alongside the \emph{Proof\ of\ Storage}. Table \ref{fig:img22222322222} shows the structure of \emph{PoW} for these nodes. 

\begin{table}[h!]
\small
\centering
\caption {PoW for nodes who are required to store blockchain segment \emph{k}}
\begin{tabular}{ccl}
\hline 
Name&Storage&Description\\\hline 
\emph{Hash\_Prev\_block}&32 bytes&The hash of the \\
&&block in block height \emph{h-s}\\
\emph{Identity\ Key}&32 bytes&Node's public identity key\\
\emph{Nonce}&32 bytes&The number tried for this PoW\\
&&to reach the required difficulty\\
\hline
\end{tabular}
\label{fig:img22222322222}
\end{table}

\begin{figure}
\centering
\begin{tikzpicture}[scale=0.84]
\node[rectangle,draw, minimum width=1.5cm, minimum height=0.7cm] (a1) at (0,1){\small {Block $i$}};
\node[rectangle,draw, minimum width=1.4cm, minimum height=0.7cm] (a2) at (4,1){\small {Block $i+1$}};
\node[rectangle,draw, minimum width=1.4cm, minimum height=0.7cm] (a5) at (8.2,1){\small {Block $i+2$}};
\node[rectangle,draw, fill=green, minimum height=0.3cm] (a3) at (1.3,1.3){\footnotesize {S}};
\node[rectangle,draw, fill=green, minimum height=0.3cm] (a4) at (1.3,.7){\footnotesize {W}};

\node[rectangle,draw, fill=red, minimum height=0.3cm] (a3) at (2,1.3){\footnotesize {S}};
\node[rectangle,draw, fill=red, minimum height=0.3cm] (a4) at (2,.7){\footnotesize {W}};
\node[rectangle,draw, dotted, minimum height=1.2cm,minimum width=1.7cm] (a44) at (1.9,1){ };
\node[rectangle,draw, fill=yellow, minimum height=0.3cm] (a3) at (2.5,1.3){\footnotesize {S}};
\node[rectangle,draw, fill=yellow,  minimum height=0.3cm] (a4) at (2.5,.7){\footnotesize {W}};
\draw (a1)--(a2);
\draw (a2)--(a5);
\draw (1.9,1.6)--(1.9,2) (1.9,2)--(8.2,2);
\draw [->](8.2,2) --(a5);
\draw (1.9,0.3)--(1.9,-0);
\node (aaa) at (4,-0.1) {\footnotesize \emph{PoW} and \emph{PoS} sent by the nodes who are required to store blockchain segment \emph{k}};
\node (aaa) at (4.7,2.15) {\footnotesize The nodes who sent \emph{PoS} and \emph{PoW} are rewarded in Block $i+2$};
\node[rectangle,draw, fill=green, minimum height=0.3cm] (a3) at (1.5+4,1.3){\footnotesize {S}};
\node[rectangle,draw, fill=green, minimum height=0.3cm] (a4) at (1.5+4,.7){\footnotesize {W}};

\node[rectangle,draw, fill=red, minimum height=0.3cm] (a3) at (2.1+4,1.3){\footnotesize {S}};
\node[rectangle,draw, fill=red, minimum height=0.3cm] (a4) at (2.1+4,.7){\footnotesize {W}};

\node[rectangle,draw, fill=yellow, minimum height=0.3cm] (a3) at (2.7+4,1.3){\footnotesize {S}};
\node[rectangle,draw, fill=yellow,  minimum height=0.3cm] (a4) at (2.7+4,.7){\footnotesize {W}};
\end{tikzpicture}
\caption{A mining procedure example. There are three occupations (red, yellow, green); \emph{S} rectangle represents \emph{Proof of Storage}; \emph{W} rectangle represents a \emph{PoW}}
\label{fig:img4}
\end{figure}
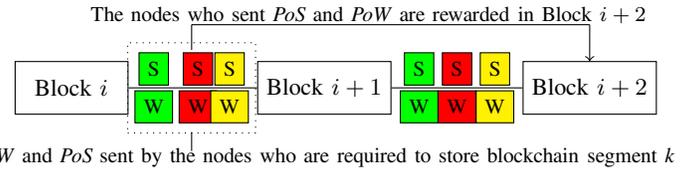

The information of the nodes which stored blockchain segment $k$ and presented \emph{Proof\ of\ Storage} and \emph{PoW} is embedded in the block of the second next block height. Figure \ref{fig:img4} shows an example of this procedure.

When a node which stores the \emph{blockchain segment} $k$,  $k=(h\ mod\ s)+1$ does not present the \emph{Proof of Storage} and the fulfilled \emph{PoW} at the block height $h$, the membership of this node is eliminated after the block height $h+2$. If this node is of occupation $X$, and there is a pending node in $PN_{X}$, then $PN_{X,1}$ replaces the eliminated node. If there is no pending node in $PN_{X}$, then $s=s-1$, the number of blocks in every blockchain segment is adjusted accordingly. After that, all nodes storing the \emph{blockchain segment} $k$ before the adjustment are back to the pending node section. In this way, we solve the challenge (4) of segment blockchain.

\subsection{Storage adjustment delay}
When $h//s$ is changed, the components of every segment is changed. Because of that, nodes need to adjust the segments they store by adding or deleting blocks, and sometimes it also needs to derive a new state from the old one. For example, in Figure 2, when $s$ changed from 2 to 3, blockchain segment two would contain "State 1" instead of "State 2". The nodes store segment two would need to derive the State 1 by acquiring segment one and execute blocks since "State 0". It is required that the nodes need to keep the old version of their segments for one block iteration after $h//s$ is changed to avoid conflicts between different versions of segments. In this way, it is providing time for nodes to change the segments smoothly.
\subsection{Reward}
There are two parts of reward in segment blockchain, one for creating a block, one for keeping the blockchain segments. The reward for creating the block is given using the same rule as Nakamoto blockchain (the reward starts from a large amount of currency at first, and cut in half in every fixed time window until reaching zero). The reward for keeping the blockchain segments is given using the following rules:
\begin{itemize}
\item When a node showed the \emph{Proof\ of\ Storage} and the fulfilled \emph{PoW} in the block height $h$, the reward is given to this node at the block of the block height $h+2$ (using the node's public identity key as the wallet address). 
\item The reward is equally divided to every node. 
\item The amount of the reward for every iteration comes from the system (as like the Nakamoto blockchain). After the reward from the system goes to zero, the reward then comes from the transaction fees.
\end{itemize}

\subsection{Power constrain}
As every node which stores the data are required to present $P\times s$ amount of difficulty per $s$ block height, the node should at least be able to generate a $P$ difficulty \emph{PoW} per iteration. We say the node which can generate $P$ difficulty \emph{PoW} in one iteration has $P$ power. Let $s=1$, and there is $n\times P$ amount of power globally. The Adversary who has $(n/2)\times P$ amount of power can only keep $n/2$ of nodes in a system of $n$ nodes because it needs to place $P$ amount of power for every Adversary node. If $s>1$, every node still needs place $P$ amount of power to maintain the spot in the system in every iteration. Otherwise, the Adversary node will be expelled at the next time window (the time window is sized $s$) for not being able to provide a $P\times s$ difficulty \emph{PoW}. If an Adversary who has $P$ amount of power stop placing power for its node $A$ and tries to gain a new node $B$, $A$ and $B$ cannot remain in the system at the same time. This is because $B$ also needs $P\times s$ of power in order to become a pending node. Thus, regardless of the number of $s$, for an Adversary who has $(n/2)\times P$ amount of power, it can only keep $n/2$ of the nodes.

\section{Combining $n/2$ blockchain sharding with Segment blockchain}
For blockchain sharding approaches, nodes only store transactions in their shards, so that the storage globally is also divided. However, since blockchain sharding approaches aim to process the transactions in different shards in parallel to improve the transaction per second globally, the system is requiring the honest nodes to be the majority of every shard. As a result, the number of shards that the transactions as a whole can be divided into, in blockchain sharding approaches, is much smaller than the number of Segments that the transactions can be divided into in Segment blockchain.

For an $n/2$ blockchain sharding system which uses Segment blockchain, the nodes keep the latest state in their shards and store different segment of the blockchain (the segments may or may not be one that contains blocks within the nodes' shards). We need to consider two failure probabilities in this system: the chance for the Adversary to control a shard and the chance for the Adversary to take all the copies of a segment. Since the two attacks are unrelated, the maximum failure probability $Pr_{max}$ for a blockchain sharding system in overall is 
\begin{equation}
    Pr_{max}=max(Pr_{{shard}_{max}},Pr_{{storage}_{max}})
\end{equation}
We know that 
\begin{equation}
    Pr_{{shard}_{max}}=(\frac{AD}{T\times S})^T
\end{equation} and 
\begin{equation}
    Pr_{{Storage}_{max}}=\frac{1}{2}^m
\end{equation}

Let
\begin{equation}
    Pr_{max}>=[(\frac{AD}{T*s_1})^T\approx\frac{1}{2}^{(n/s_0)}]
    \label{eq:111}
\end{equation} where  $0.5*n/s_1<T<=n/s_1$,  \emph{T} is a pre-defined setting \footnote{The smaller the \emph{T} is, the less secure a shard is, the larger the \emph{T} is, the easier a shard can be halted by the Adversary \cite {XUXU}.}

We can use equation \ref{eq:111} to calculate the maximum number of shards $s_1$ required for the n/2 blockchain sharding approach to function securely and the maximum number of segments $s_0$ the Segment blockchain embedded to the system can have in order to maintain the same security threshold $Pr_{max}$.

Let $AD=\frac{n}{2}$ (this is the maximum $AD$, because if exceed this number the Adversary would be the majority). Then, we can derive:
\begin{equation}
    s_0=-\frac{n\times log(2)}{log(2^{-T}\times(\frac{n}{s_1\times T})^T)}
\end{equation}

\begin{figure}[h!]
\centering
\includegraphics[width=0.5\textwidth]{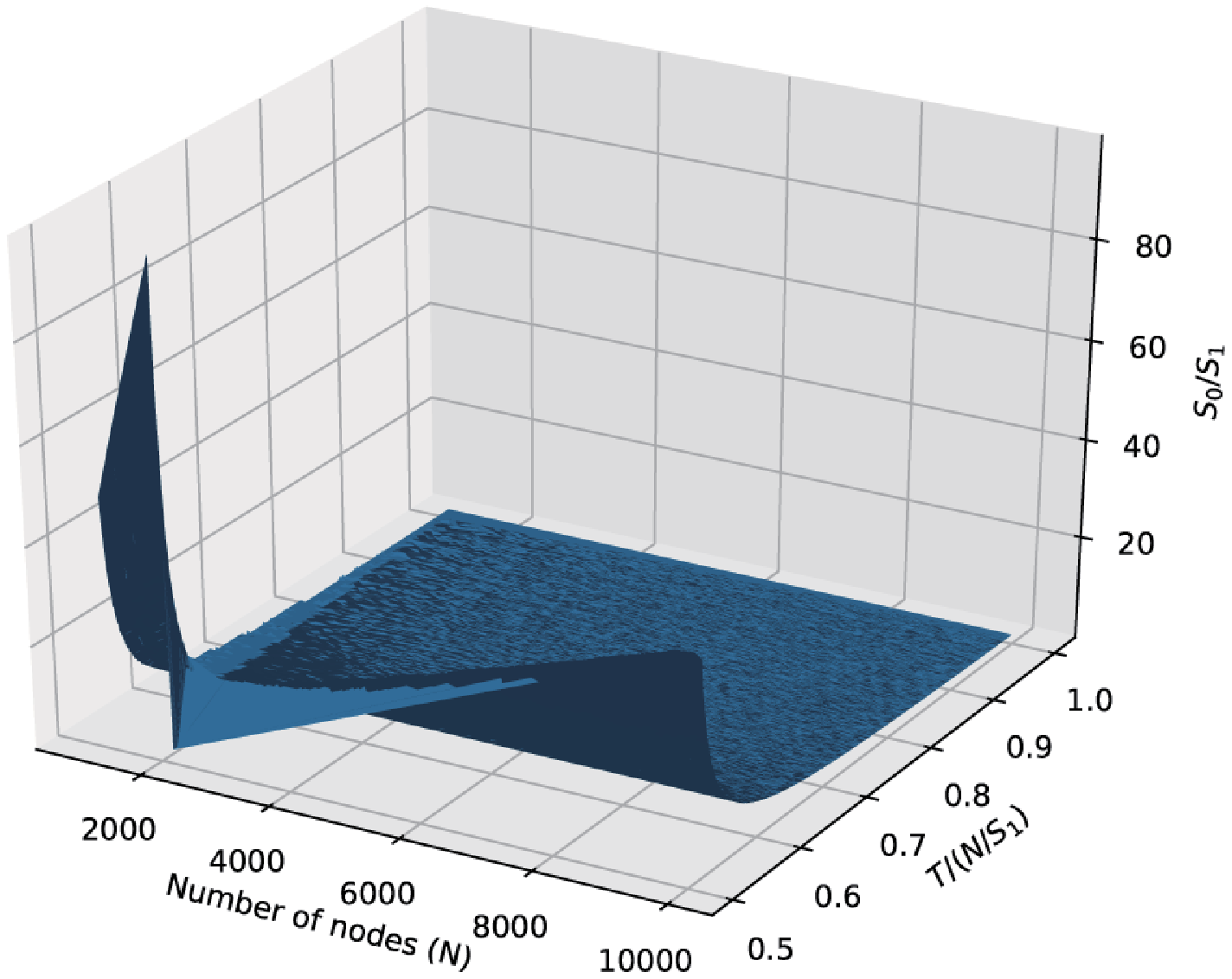}
\caption{$s_0/s_1$}
\label{fig:img7717}
\end{figure}

Assumed we set the maximum failure probability to be $Pr_{max}=10^{-6}$ (this bounds the number of $s_1$), then we derive $\frac{S_0}{S_1}$ which is shown in Figure \ref{fig:img7717}.

As the data in Figure \ref{fig:img7717} suggested, the blockchain can be divided into $\frac{S_0}{S_1}$ more times of segments than shards, so that the nodes can store a much smaller number of blocks than store all the blocks of a shard. Thus, if a blockchain sharding system separates the transaction storage from transaction verification by embedding a segment blockchain, the storage requirement for an individual node can be reduced significantly. 

\section{Data requirement}
Let a record in \emph{State} sized $41\ bytes$ (a wallet address sized $33\ bytes$, the balance of a wallet address sized $8\ bytes$). Let a block of the Nakamoto blockchain sized $SB\ bytes$, a block of segment blockchain sized $SB + Size_{Pending\ node\ section}$. A record in $Pending\ node\ section$ sized $68 bytes$ ($4\ bytes$ for the occupation of this node, $32\ bytes$ for the public identity key and $32\ bytes$ for the \emph{PoW} it demonstrated). In reality, we can use persistent data structure \cite{kurosawa1996method} to store \emph{States}. In this way, the size of storage can be further reduced.

\begin{figure}[h!]
\centering
\includegraphics[width=0.5\textwidth]{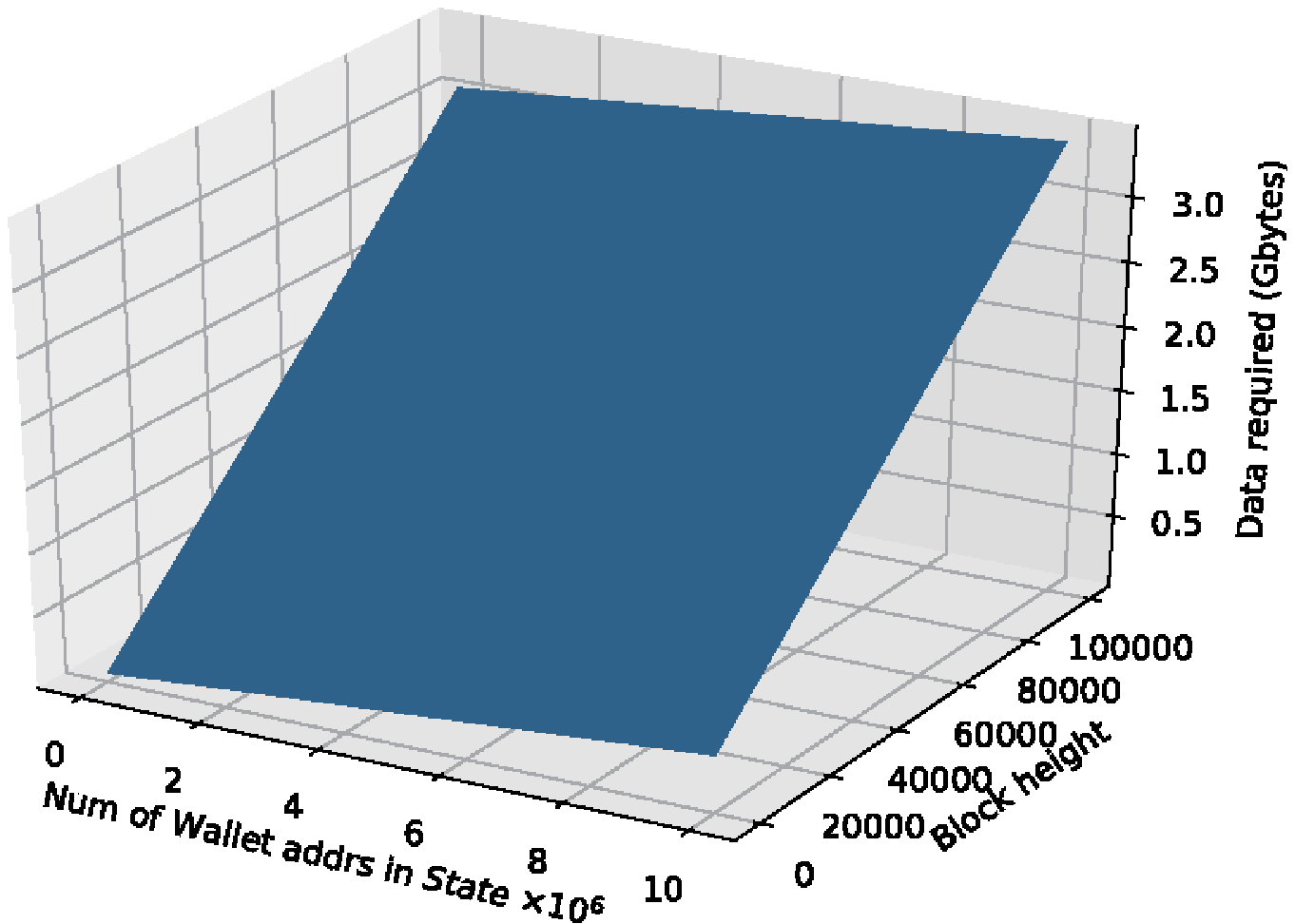}
\caption{Segment blockchain data requirement}
\label{fig:img555}
\end{figure}

\begin{figure}[h!]
\centering
\includegraphics[width=0.5\textwidth]{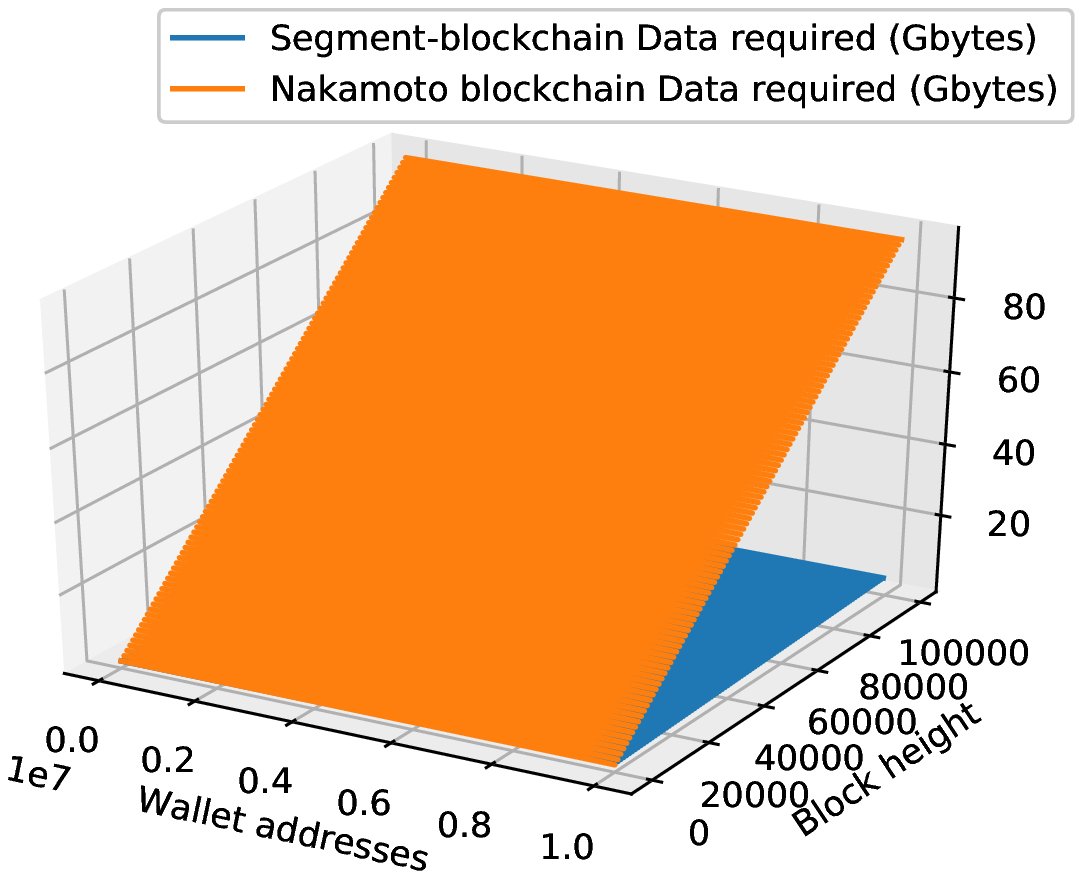}
\caption{Segment blockchain data requirement compared to Nakamoto blockchain data requirement}
\label{fig:img777}
\end{figure}

Figure \ref{fig:img555} shows the data requirement for the segment blockchain with different number of records in the \emph{State} (let $SB=1\ Mbytes$, $m=256$), and with the changes of the block height \emph{h}. Let there be $8000$ nodes (the number of the Bitcoin nodes currently), $256$ pending nodes are in the pending node section in every block. Figure \ref{fig:img777} shows the differences in the amount of data stored in a node in the Nakamoto blockchain and the amount of data stored by a node in segment blockchain with the different number of accounts in the \emph{state}. Segment blockchain largely shrank the data required while maintaining the full functions of Nakamoto blockchain.

As the analysis in Section 2 showed, the chance for the Adversary who has $50\%$ of the overall calculation power to store all the copies of a block is ${\frac{1}{2}}^{m}$. When $m=256$, the security of segment blockchain reached the security level of the standard public-private encryption systems. It is very safe to use an $m=256$ segment blockchain to power financial systems because the encryption systems of the same security threshold are already widely tested by the public and used in many online banking systems.

\section{Conclusion}
In this paper, we showed an approach to reduce the storage requirement of the blockchain system while keeping the decentralisation without compromising the security of the blockchain. The data analyses proved that segment blockchain largely reduced the data requirement compared to Nakamoto blockchain. Thus, it is of more advantage to using segment blockchain to power data-heavy blockchains.




%
{
\bibliographystyle{unsrt}
\bibliography{sample}}
\end{document}